\documentclass[11pt,a4paper]{article}
\usepackage{jheppub}

\usepackage[english]{babel}

\pdfoutput=1 
\RequirePackage{ifpdf}
\usepackage{amsmath}
\usepackage{mathtools}
\usepackage{amssymb}
\usepackage{braket}

\usepackage{hyperref}

\usepackage{notoccite}

\usepackage{amsfonts}
\usepackage{float}

\usepackage{multirow}
\usepackage{graphicx}

\usepackage[normalem]{ulem}

\newcommand\as{\alpha_{\mathrm{S}}}

\title{Soft contributions to heavy quark production in arbitrary kinematics}

\author[a]{Simone Devoto,}
\author[b]{Javier Mazzitelli}

\affiliation[a]{Department of Physics and Astronomy, Ghent University, 9000 Ghent, Belgium}
\affiliation[b]{Paul Scherrer Institut, CH-5232 Villigen PSI, Switzerland}

\emailAdd{simone.devoto@ugent.be}
\emailAdd{javier.mazzitelli@psi.ch}

\abstract{
We present the computation of the soft-parton contribution at low transverse momentum for the production of a heavy quark pair in association with a colour singlet at the next-to-next-to-leading order (NNLO) in the QCD coupling.
This paper extends to arbitrary kinematics previous results obtained under the assumption of a back-to-back configuration for the final-state emitters.
These new results take into account the production of an additional colour singlet system, thus allowing for the evaluation of the resummation formula for associated heavy-quark pair production at NNLO, and the implementation of the $q_T$ subtraction formalism for this class of processes.
We provide the results in the form of a code for their on-the-fly evaluation.
}

\preprint{}

\begin{document}
\allowdisplaybreaks[4]
\unitlength1cm
\maketitle
\flushbottom

%%%%%%%%%%%%%%%%%%%%%%%%%%
\section{Introduction}
\label{sec:intro}

The production of a heavy-quark pair associated with a colour singlet plays a central role in the precision physics programme at the LHC.
The measurement of a top-antitop quark pair in association with a Higgs boson ($t\bar tH$)~\cite{ATLAS:2018mme, CMS:2018uxb, ATLAS:2021qou, CMS:2024fdo, ATLAS:2024gth} provides us with a powerful tool to study the Higgs boson properties and a direct access to the top-quark Yukawa coupling.
The observation of the associated production of a top quark pair with a gauge boson ($t\bar tZ$, $t\bar t W$, $t\bar t \gamma$)~\cite{ATLAS:2015qtq, CMS:2015uvn, ATLAS:2016wgc, CMS:2017ugv, ATLAS:2019fwo, CMS:2021ugv, CMS:2022tkv, ATLAS:2023eld, ATLAS:2024moy, CMS:2022lmh, ATLAS:2020yrp} offers us additional information to study the Standard Model, both as a background for different measurements (such as $t\bar t H$ production) and as a direct signal to probe the EW couplings of the top quark.
The associated production with a $W$ boson, in particular, presents an ongoing tension between theoretical prediction and experimental measurement.
In addition, the associated production of lighter quark flavours, mainly bottom quarks, can also be regarded as associated heavy-quark production, at least in certain kinematical regions, and has received the attention of the LHC collaborations~\cite{ATLAS:2011jbx,ATLAS:2013gjg,CMS:2013xis,CMS:2016eha,CMS:2021pcj,ATLAS:2024tnr}.

An accurate theoretical description of this class of processes is a high priority for both Standard Model measurements and new physics searches, and attracted the efforts of the community in the last decades.
At the dawn of the High-Luminosity phase of the LHC, the inclusion of next-to-next-to-leading (NNLO) QCD corrections for associated heavy-quark production becomes of utmost importance.
First NNLO QCD corrections to associated heavy-quark production have been recently carried out by applying different approximations to the two-loop amplitudes and estimating their impact, since exact two-loop amplitudes for these processes are currently unavailable. This provided first predictions at NNLO accuracy for $t\bar tH$~\cite{Catani:2022mfv, Devoto:2024nhl}, and $t\bar t W$~\cite{Buonocore:2023ljm} production. 
These computations, as well as a prediction for $b\bar b W$ production including massive bottoms in the final state~\cite{Buonocore:2022pqq} and developements towards an NNLO description for $b\bar b W^+W^-$~\cite{Buonocore:2025fqs}, have been obtained by using the $q_T$-subtraction formalism~\cite{Catani:2007vq} in order to handle and cancel infrared (IR) divergences arising at intermediate steps of the computation.

The $q_T$ subtraction formalism uses our knowledge on the IR behaviour of the cross-section in the $q_T\to0$ limit given by resummation studies to construct a counterterm. 
Originally developed for the production of a colourless final state, results from resummation studies for heavy-quark production~\cite{Zhu:2012ts,Li:2013mia,Catani:2014qha,Catani:2018mei,Catani:2017tuc,Ju:2022wia} made its generalisation to this class of processes possible~\cite{Catani:2019iny, Catani:2019hip,Catani:2020tko,Catani:2020kkl}.
The resummation formalism was extended to the production of heavy quarks in association with a colour singlet in Ref.~\cite{Catani:2021cbl}, where all the coefficients needed for the expansion of the resummation formula up to NLO were given.
The non-zero mass of the final-state quarks regulates any possible collinear divergence, but allows for additional soft divergences not included in the colourless case.
These soft-parton contributions have been computed in Ref.~\cite{Catani:2023tby}  in the specific case of a back-to-back kinematics for the heavy quarks in the final state (see also Ref.~\cite{Angeles-Martinez:2018mqh} for an independent calculation in the context of soft collinear effective theory).
The phenomenological results in Refs.~\cite{Catani:2022mfv, Devoto:2024nhl,Buonocore:2023ljm, Buonocore:2022pqq,Buonocore:2025fqs} have been obtained building upon this framework, but the more involved kinematics of 5-points processes required to drop the assumption of the back-to-back configuration of Ref.~\cite{Catani:2023tby} in the soft contributions. This extension is the topic of this paper.

The soft-parton contributions presented in Ref.~\cite{Catani:2023tby} are also a necessary ingredient in the extension of the {\sc{MiNNLO}}$_\text{PS}$~\cite{Monni:2019whf,Monni:2020nks} method to processes with heavy quarks, and they enter the NNLO+PS event generator for $t\bar{t}$ production presented in Refs.~\cite{Mazzitelli:2020jio,Mazzitelli:2021mmm}, as well as the $b\bar{b}$ generator of Ref.~\cite{Mazzitelli:2023znt}. In an analogous way to what happens for the $q_T$-subtraction method, in order to obtain an NNLO+PS generator for associated heavy quark production within the {\sc{MiNNLO}}$_\text{PS}$ method, the extension of the soft-parton contributions to general kinematics was needed. The first NNLO+PS applications for these type of processes have already appeared in the literature, for $b\bar{b}Z$~\cite{Mazzitelli:2024ura} and $b\bar{b}H$~\cite{Biello:2024pgo} production.

In this paper, we report on the computation of such soft-parton contributions in an arbitrary kinematical configuration. 
Our results are provided in the form of a C++ library which can compute them on-the-fly in an arbitrary phase-space point.
This paper is structured as follows. In section~\ref{sec:int} we introduce the soft integrals entering our calculation and how they extend the results of Ref.~\cite{Catani:2023tby}. In section~\ref{sec:results} we briefly describe our methodology for their evaluation and we provide results in some benchmark points. We also present \textsc{Shark}, the C++ library for the on-the-fly evaluation of the soft-parton contributions and finally, in section~\ref{sec:conclusions}, we summarise our results.

%%%%%%%%%%%%%%%%%%%%%%%%%%%%%%%%%%%%%%%%%%%%%%%%%%%%%%%%%%%%%%%%%%%%%%%%%%%%%%%%%%%%%

\section{Soft Integrals}
\label{sec:int}

We consider, at the partonic level, the associated production of a heavy-quark pair:
\begin{equation}
\label{eq:process}
a_1(p_1)+a_2(p_2)\to Q(p_3)+ \bar{Q}(p_4)+F(p_5)+X\;.
\end{equation}
The symbol $a_i=q, \bar q, g$ represents any of the possible initial-state massless partons, while $Q$ stands for the final-state heavy quark $Q$ of mass $m$.
$F$ is a colour singlet system carrying a total momentum $p_5$, while $X$ represents the additional final-state radiation.
The kinematics of the final state $Q\bar Q F$ is entirely specified by the total momentum $q=p_3+p_4+p_5$, with $q^2=M^2$ being the invariant mass of the final state, and a set of angular variables $\vec \Omega$.
We indicate with $q_T$ the total transverse momentum of the final-state: $q_T=p_{3T}+p_{4T}+p_{5T}$. 

Eq.~(\ref{eq:process}) can describe a wide class of processes, but they are all share a common IR structure~\cite{Catani:2000ef,Mitov:2009sv,Mitov:2010xw,Ferroglia:2009ep,Ferroglia:2009ii}.
While at LO the absence of extra radiation enforces $q_T=0$, higher perturbative orders introduce a $q_T$ dependence to the cross-section. Because of the presence of large logarithmic contributions of the form $\as^{n+2}\frac{1}{q_T^2} \ln^k (M^2/q_T^2)$, a proper description of the $q_T\to0$ limit requires their resummation at all orders. In this framework, the cross-section for the production of the final state $Q\bar Q F$ from the collision of two hadrons $c_1(P_1)$, $c_2(P_2)$ can be written , starting from the LO contribution $\left[ d\sigma^{(0)}_{c_1 c_2}\right]=\as^2(M^2)/M^2d{\hat \sigma}^{(0)}_{c_1c_2\to Q\bar QF}/d{\vec \Omega}$, as~\cite{Catani:2014qha}:
\begin{align}
\label{eq:resumm}
    &\frac{d\sigma(P_1,P_2;\, \vec q_T, M, y, \vec \Omega)}{d^2\vec q_T\, dM^2\, dy\, d\vec\Omega}=\frac{M^2}{2P_1\cdot P_2}\sum_{c\bar c}\left[ d\sigma^{(0)}_{c_1 c_2}\right]\int\frac{d^2\vec b}{(2\pi)^2}e^{i\vec b\cdot \vec q_T}S_c(M,b)
    \nonumber\\
    &\times \sum_{a_1, a_2}\int_{x_1}^{1} \frac{dz_1}{z_1}\int_{x_2}^1\frac{dz_2}{z_2} [(\mathbf{H\Delta})C_1C_2]_{c\bar c;a_1a_2} f_{a_1/h_1}(x_1/z_1,b_0^2/b^2)f_{a_2/h_2}(x_2/z_2,b_0^2/b^2)\;.
\end{align}
Here, the integration over $z_1$, $z_2$ performs the convolution of the partonic cross-section with the parton distribution functions $f_{a/h}(x,\mu_F^2)$, with the lower limits of integration being 
\begin{equation}
    x_{1,2}=\frac M{\sqrt{2P_1\cdot P_2}}\, e^{\pm y}\;,
\end{equation}
where $y$ is the rapidity of the final state $Q\bar QF$: $y=1/2 \ln (q\cdot P_2/q\cdot P_1)$.
The functions $C_i$, $S_c$ and $\mathbf{H\Delta}$ are perturbative ingredients that encapsulate the behaviour of additional radiation in different phase-space regions. 
Their computation is performed in the impact parameter ($\vec b$) space, which allows for a factorisation of the kinematics of the multi-particle emissions.
The functions $C_i$ contain contributions coming from emissions collinear to the initial-state partons $a_i$ in the small momentum region $q\lesssim 1/b$. Meanwhile, the Sudakov form factor $S_c$ includes flavour-conserving emissions in the soft-collinear regions, at scales  $1/b\lesssim q\lesssim M$.
Both $C_i$ and $S_c$ are process-independent: $\mathbf{H\Delta}$ contains instead the information on  process-dependent corrections, including the soft radiation from the $Q\bar QF$ final state.
This contribution has been expressed in Ref.~\cite{Catani:2023tby} in terms of the colour space operator $\mathbf{h}(\as)$, which has been computed therein in the specific case of heavy-quark pair production, thus imposing the constraint of a back-to-back configuration of final-state emitters.
In this paper we remove this constraint, and we present a result for $\mathbf{h}$ valid in arbitrary kinematics.
In the following, we quickly summarise the definition of $\mathbf{h}$ and the main ingredients for its evaluation, referring the interested reader to Ref.~\cite{Catani:2023tby} for a more comprehensive discussion.

The operator $\mathbf{h}$ can be defined as:
\begin{align}
    \label{eq:h}
    &\mathbf{h}(\as)=
    1+\frac{\as}{2\pi}\braket{\mathbf{F}_{{\rm ex},1}^{(0)}}_{\rm av.}
    +\left(\frac{\as}{2\pi}\right)^2\Big\{\braket{(\mathbf{F}_{{\rm ex},1}^{(0)})^2}_{\rm av.}-\frac 12 \left(\braket{\mathbf{F}_{{\rm ex},1}^{(0)}}_{\rm av.}\right)^2
    \nonumber\\
    &\phantom{\mathbf{h}(\as)=}+\braket{\mathbf{F}_{{\rm ex},2}^{(0)}}_{\rm av.}-2\pi\beta_0\braket{\mathbf{F}_{{\rm ex},1}^{(1)}}_{\rm av.}-\frac 14\left[\left(\mathbf{\Gamma}^{(1)}_{\text{sub}}-{\rm h.c.}\right),\langle\mathbf{F}_{{\rm ex},1}^{(1)}\rangle_{\rm av.}\right]      \Big\}+\mathcal{O}(\as^3)\;,
\end{align}
where $\mathbf{\Gamma}_\text{sub}$ is the subtracted soft anomalous dimension  introduced in Ref.~\cite{Catani:2014qha} and the brackets denote the average with respect to the azimuthal degrees of freedom $\phi$.
In order to regularize both ultraviolet and IR divergences, we use dimensional regularisation, performing our calculations in $D=4-2\epsilon$ dimensions. The operators $\mathbf{F}^{(n)}_{{\rm ex}, m}(\phi)$ are the coefficients of the Taylor expansion of the operator $\mathbf{F}_{{\rm ex}}(\vec b)$ in the parameter $\epsilon$:
\begin{equation}
    \mathbf{F}_{{\rm ex}}(\vec b)=\sum_{m,n}\left(\frac{\alpha_0}{2\pi} S_\epsilon \right)^m\,\left(\frac{b^2\, \mu_0^2\, e^{-2\gamma_E}}{4}\right)^{m \epsilon} \, \mathbf{F}^{(n)}_{{\rm ex}, m}(\phi)\,\epsilon^n\;.
\end{equation}
$\mathbf{F}_{{\rm ex}}(\vec b)$ is expanded in the bare QCD coupling $\alpha_0$. 
The relation between the bare coupling and the running coupling is the following:
\begin{equation}
	\label{eq:renorm}
	\alpha_0\mu_0^{2\epsilon} S_\epsilon=\as(\mu_R^2) \mu_R^{2\epsilon}\left(1-\as(\mu_R^2) \frac{\beta_0}{\epsilon}
        +\mathcal O(\as^2)\right)\;,
\end{equation}
with
\begin{equation}
  12\pi \beta_0=11 C_A -2 n_f\;,\hspace{1cm}S_\epsilon=(4\pi)^\epsilon e^{-\epsilon \gamma_E}\;,
\end{equation}
and $\gamma_E$ being the Euler number.

The coefficients $\mathbf{F}^{(n)}_{{\rm ex}, m}(\phi)$ can be obtained from the integration of the soft parton contributions at the required perturbative order, after the subtraction of the corresponding contribution originated from initial-state emission.
At NLO we have:
\begin{align}
 \label{eq:Fex1I} 
 & 2\times\frac{S_\epsilon}{8\pi^2}\left(\frac{b^2}{b_0^2}\right)^\epsilon \mathbf{F}_{{\rm ex},1}(\phi)=\mathbf{I}_g^{(0)}(\vec{b})\, ,\\
 \label{eq:I0g}
 & \mathbf{I}^{(0)}_g(\vec{b})=-\int \frac{d^Dk}{(2\pi)^{D-1}}\delta_+(k^2)\left|\mathbf{J}^{(0)}_g(k)\right|^2_{\rm sub} e^{i \vec{b}\cdot \vec k_T}\;,
\end{align}
where $\left|\mathbf{J}_g^{(0)}(k)\right|^2_{\text{sub}}$ is the square of the tree-level soft current $\mathbf{J}^{(0)}_{g,\mu}$
\begin{equation}
    \label{eq:nlo current}
  	\mathbf{J}^{(0)}_{g,\mu}(k)=\sum_{i=1}^4 \textbf{T}_i\frac {p_{i\mu}}{(p_i\cdot k)}
\end{equation}
after subtracting the contribution of the initial-state emission:
\begin{align}
  \label{eq:single_gluon_NLO}
  \left|\mathbf{J}_g^{(0)}(k)\right|^2_{\text{sub}}=& \left|\mathbf{J}_g^{(0)}(k)\right|^2- \left|\mathbf{J}_g^{(0)}(k)\right|^2_{\rm colourless}\nonumber\\
  =&\sum_{j=3,4}\Bigg[\frac{m^2}{(p_j\cdot k)^2}\textbf{T}_j^2
  +2\sum_{i=1,2}\left(\frac{p_i\cdot p_j}{p_j\cdot k}-\frac{p_1\cdot p_2}{(p_1+p_2)k}\right)\frac{\textbf{T}_i\cdot\textbf{T}_j}{p_i\cdot k}\Bigg]
  \nonumber\\&
  +\frac {2p_3\cdot p_4}{(p_3\cdot k) (p_4\cdot k)}\textbf{T}_3\cdot\textbf{T}_4
  \;.
\end{align}
The corresponding expressions at NNLO read:
\begin{align}
& 2\times\frac{S_\epsilon^2}{(8\pi^2)^2}\left(\frac{b^2}{b_0^2}\right)^{2\epsilon} \mathbf{F}_{{\rm ex},2}(\phi)=\mathbf{I}_g^{(1)}(\vec{b})+\mathbf{I}_{q\bar{q}}^{(0)}(\vec{b})+\mathbf{I}_{gg}^{(0)}(\vec{b})\;,\\
  &\mathbf{I}^{(1)}_g(\vec{b})=-\int \frac{d^Dk}{(2\pi)^{D-1}}\delta_+(k^2)\left(\mathbf{J}^{(0)\dagger}_g(k)\mathbf{J}^{(1)}_g(k)+{\rm c.c.}\right)_{\rm sub} e^{i \vec{b}\cdot \vec k_T} \label{eq:Ibom1g}\;,\\
  &\mathbf{I}^{(0)}_{q\bar{q}}(\vec{b})=\int \frac{d^Dk_1}{(2\pi)^{D-1}} \frac{d^Dk_2}{(2\pi)^{D-1}}\delta_+(k_1^2)\delta_+(k_2^2)
  {\boldsymbol I}^{(0)}_{q\bar{q}}(k_1,k_2)\big|_{\rm sub} e^{i \vec{b}\cdot (\vec k_{T1}+k_{T2})}\label{eq:Ibomqq}\;,\\
&\mathbf{I}^{(0)}_{gg}(\vec{b})=\frac{1}{2}\int \frac{d^Dk_1}{(2\pi)^{D-1}} \frac{d^Dk_2}{(2\pi)^{D-1}}\delta_+(k_1^2)\delta_+(k_2^2)  \mathbf{W}^{(0)}_{gg}(k_1,k_2)\big|_{\rm sub} e^{i \vec{b}\cdot (\vec k_{T1}+k_{T2})}\label{eq:Ibomgg}\;,
  \end{align}
with the following definition of the soft factors $\left|\mathbf{J}^{(0)}_g(k)\right|^2_{\rm sub}$, $\left(\mathbf{J}^{(0)\dagger}_g(k)\mathbf{J}^{(1)}_g(k)+{\rm c.c.}\right)$, $\boldsymbol{I}^{(0)}_{q\bar{q}}(k_1,k_2)$ and $\mathbf{W}^{(0)}_{gg}(k_1,k_2)$:
\begin{align}
  \label{eq:J0J1pcc}
  &\mathbf{J}_g^{(0)\dagger}(k)\cdot \mathbf{J}_g^{(1)}(k)+{\rm c.c.}=2C_A\sum_{i\neq j}\left(\frac{(p_i\cdot p_j)}{(p_i\cdot k)(p_j\cdot k)}-\frac{m^2}{(p_j\cdot k)^2}\right){\cal R}_{ij}\, \mathbf{T}_i\cdot \mathbf{T}_j\nonumber\\
    &\phantom{\mathbf{J}_g^{(0)\dagger}(k)\cdot \mathbf{J}_g^{(1)}(k)+{\rm c.c.}=}-4\pi\sum_{i,j,k}{\vphantom{\sum}}'
    \frac{p_i\cdot p_j}{(p_i\cdot k)(p_j\cdot k)} {\cal I}_{ik} f^{abc} T_i^a T_k^bT^c_j\, ,
\\
\label{eq:taskqq_2}
&{\boldsymbol I}_{q\bar q}^{(0)}(k_1,k_2)=\frac{T_R}{(k_1\cdot k_2)^2}\left[\mathbf{J}_{g,\mu}^{(0)}(k_1+k_2)\right]^\dagger \left(-g^{\mu\nu} k_1\cdot k_2 +k_1^\mu k_2^\nu+k_1^\nu k_2^\mu\right)\,\mathbf{J}^{(0)}_{g,\nu}(k_1+k_2)+...\;,
\\&
\label{eq:ggW}
\mathbf{W}^{(0)}_{gg}(k_1,k_2)=-C_A \sum_{i,j=1}^n \mathbf{T}_i \cdot \mathbf{T}_j \; {\cal S}_{ij}(k_1,k_2)\, .
\end{align}
The expression for the functions ${\cal R}_{ij}$, ${\cal I}_{ij}$ in Eq.~(\ref{eq:J0J1pcc}) can be found in Ref.~\cite{Bierenbaum:2011gg} and, in a simplified form for the case of two massive emitters in Ref.~\cite{Czakon:2018iev}, while the soft factor ${\cal S}_{ij}$ is provided in Ref.~\cite{Catani:2023tby}.
Eq.~(\ref{eq:taskqq_2}) drops some gauge-dependent pieces which are proportional to the total colour-charge of the hard partons, and thus not contributing to the case at hand.

%%%%%%%%%%%%%%%%%%%%%%%%%%%%%%%%%%%%%%%%%%%%%%%%%%%%%%%%%%%%%%%%%%%%%%%%%%%%%%%%%%%%%

\section{Results}
\label{sec:results}

We start by pointing out the main differences with respect to the results presented in Ref.~\cite{Catani:2023tby} for the case of $Q\bar{Q}$ production. Although there are no conceptual differences in the IR structure of the scattering amplitudes, the removal of the back-to-back constraint for the heavy quarks, valid in the $Q\bar{Q}$ Born kinematics, poses a significant challenge in the calculation of the soft integrals at hand. This is only true, however, for the contributions that depend simultaneously on both heavy-quark momenta, and therefore many of the results from Ref.~\cite{Catani:2023tby}, namely those depending on the momentum of only one of the heavy quarks, can be recycled for the $Q\bar{Q}F$ case. More specifically, the contributions that had to be recomputed are: 
(i) the ${\cal R}_{34}$ term in the one-loop soft gluon emission from Eq.~(\ref{eq:J0J1pcc}), 
(ii) the term proportional to $\mathbf{T}_3 \cdot \mathbf{T}_4$ in the soft-quark-pair production term from Eq.~(\ref{eq:taskqq_2}),
(iii) the ${\cal S}_{34}$ term in the double-gluon soft current from Eq.~(\ref{eq:ggW}) and 
(iv) the contributions proportional to the azimuthal average of the square of the NLO result, i.e. the $\braket{(\mathbf{F}_{{\rm ex},1}^{(0)})^2}_{\rm av.}$ term in Eq.~(\ref{eq:h}).
In addition, the contributions proportional to the colour structure $f^{abc} T_i^a T_k^bT^c_j$, which gave a vanishing contribution upon contraction against the $Q\bar{Q}$ matrix element, had to be included in the general $Q\bar{Q}F$ case for gluon-initiated processes.
This includes the ${\cal I}_{ik}$ terms in the one-loop soft current from Eq.~(\ref{eq:J0J1pcc}) and the commutator between the non-hermitian part of the soft anomalous dimension and the ${\cal O}(\epsilon)$ of the NLO contribution to $\mathbf{h}$, see last term in Eq.~(\ref{eq:h}).

While some of these ingredients were already computed numerically for $Q\bar{Q}$ production, the results in Ref.~\cite{Catani:2023tby} for such contributions were pre-computed and implemented as two dimensional grids, which were then used to perform a splines fit in order to provide a flexible implementation. In the case of $Q\bar{Q}F$ production, we decided to change the approach and perform an on-the-fly calculation of such numerical pieces. This is motivated by the fact that the larger dimension of the Born phase space would pose a challenge for an interpolation procedure.

Before going into the details of the calculation for the different contributions, we make a general remark regarding the computation of the azimuthal average.
Since we work in $D=4-2\epsilon$ dimensions, the transverse component of the soft-emitted particles, $\vec{k}_T$, is a $D-2$ dimensional vector, which in general can be described with $D-3$ angles.
The back-to-back kinematics analysed in Ref.~\cite{Catani:2023tby} implied that the soft integrals could depend only on one non-trivial azimuthal degree of freedom, nominally the angle  $\phi_1$ between the top-quark and the soft radiation, and this was therefore the only non-trivial azimuthal average to be performed.
In the present case, the absence of the back-to-back constraint implies that the top and anti-top transverse momenta can be parametrized as
\begin{align}
    \vec{p}_{T,3} &= p_{T,3} (1,0,\dots,0) \\
    \vec{p}_{T,4} &= p_{T,4} (\cos\chi,\sin\chi,\dots,0)     
\end{align}
with $\chi$ being the angle between the top and antitop quark in the transverse plane, namely $\cos\chi = (\vec{p}_{T,3} \cdot \vec{p}_{T,4} )/(p_{T,3} p_{T,4})$, and where without loss of generality we have chosen the reference frame such that $\vec{p}_{T,3}$ and $\vec{p}_{T,4}$ have only one and two non-zero components, respectively.
Therefore, in any integrand depending simultaneously on $\vec{p}_{T,3}$ and $\vec{p}_{T,4}$, the product against the soft momentum $\vec{k}_T$, generically written as
\begin{align}
        \vec{k}_{T} &= k_{T} (\cos\phi_1,\sin\phi_1 \cos\phi_2,\dots)   
\end{align}
will lead to an additional dependence on $\phi_2$. We note that this dependence would vanish in the case $\sin\chi = 0$, as it happens in the back-to-back limit.
The ensuing azimuthal average of a function $F(\phi_1, \phi_2)$ thus now takes the form:
\begin{equation}
	\label{eq:azimuthal average}
	\braket{F(\phi_1,\phi_2)}_{\rm av.}=-\frac\epsilon\pi
    \int_0^\pi d\phi_1 d\phi_2 \sin^{-2\epsilon}\phi_1 \sin^{-1-2\epsilon}\phi_2 F(\phi_1,\phi_2)\;.
\end{equation}
Note that the term $\sin^{-1-2\epsilon} \phi_2$ is singular in the $\epsilon\to0$ limit.
As a consequence, its expansion in powers of $\epsilon$ needs to be expressed in terms of distributions.

We finally note that in Ref.~\cite{Catani:2023tby} the computation has been performed in the Fourier ($b$-space) representation of the soft integrals, related to the direct ($q_T$-space) representation by the following formal substitution in the integrals:
\begin{equation}
\label{eq:phasespace}
\delta^{(D-2)}\left(\vec q_T+ \vec k_{T}\right)~~~\longrightarrow~~~e^{i \vec{b}\cdot\vec k_{T}}\, .
\end{equation}
This was found to be convenient in order to achieve more compact analytical expressions for the intermediate results.
In the present computation, however, we found the $q_T$-space more suitable for a numerical integration, since the presence of the Dirac delta makes the integration over $D-2$ dimensions of the soft momenta exchanged $k$ trivial.

\subsection{One-loop soft gluon emission}
Among the new pieces that had to be computed, the numerical integration of the one-loop soft contribution $\mathbf{I}^{(1)}_g$ is the simplest one, given that we only need to integrate over the phase space of one soft gluon. 
After imposing the $\vec k_{T} = -\vec q_T$ condition, one of the two remaining integrals over $d^Dk$ can be trivially performed thanks to the on-shell condition, imposed by the $\delta_+(k^2)$. Therefore, before azimuthal average, only a one-fold integration remains. 
The integration measure takes thus the following form:

\begin{align}
\label{eq:1gstructure}
    & \int d^Dk\;\delta_+(k^2) \;\delta^{(D-2)}(\vec q_T+ \vec k_{T}) 
    =\int_0^\infty \frac {dy}{2y}
    \;,
\end{align}
where, introducing the light-cone coordinates $p_\pm=\frac{p_0\pm p_z}{\sqrt2}$, the scalar products $p_i\cdot k$ can be written in terms of the integration variable $y = k_-/q_T$ as: 
\begin{align}
    p_i\cdot k=&q_T\left(p_{i+}y+\frac{p_{i-}}{2y}\right)+\vec p_{iT}\cdot\vec q_T\;.
\end{align}
The angular dependence of the integrand in Eq.~(\ref{eq:1gstructure}) is then contained in the scalar products $\vec p_{iT}\cdot \vec k_T$:
\begin{align}
    & \vec p_{3T}\cdot \vec k_T= p_{3T} k_T \cos\phi_1\;,\\
    & \vec p_{4T}\cdot \vec k_T= p_{4T} k_T(\cos\phi_1\cos\chi+\sin\phi_1\cos\phi_2\sin\chi)\;.
\end{align}
Performing the azimuthal average of Eq.~(\ref{eq:1gstructure}) leads to two and three-fold integrations that were performed numerically with a Monte Carlo method.

\subsection{Soft quark-pair emission}
\label{sec:qq}
The contribution to the soft current from the emission of a soft quark pair from the final-state emitters is contained in the integral $I^{q\bar q}_{34}$:
\begin{equation}
    \mathbf{I}^{(0)}_{q\bar q}({\vec b})=-\frac{F(\epsilon)}{(2\pi)^{D-1}}\left\{\sum_{j=3,4} \left[I^{q\bar q}_{jj}({\vec b})\,\mathbf{T}_j^2+2\sum_{i=1,2} I^{q\bar q}_{ij}({\vec b})\,\mathbf{T}_i\cdot\mathbf{T}_j\right]+2I^{q\bar q}_{34}({\vec b})\,\mathbf{T}_3\cdot \mathbf{T}_4\right\}\;,
\end{equation}
which is expressed in Ref.~\cite{Catani:2023tby} in terms of the integral functions $L_n$, $P_n$:
\begin{align}
    \label{eq:iqq34_result}
    &\langle I^{q\bar q}_{34}(\vec b)\rangle_{\rm av.}= \pi ^{1-\epsilon } \Gamma (1-\epsilon ) \Gamma (-2 \epsilon )\left(\frac{b^2}{4}\right)^{2\epsilon} \frac{1+\beta^2}{2\beta}
\left\{-\frac{1}{\epsilon} L_0 - 2 L_1+ \epsilon ( 2 P_2 - L_2 )+ {\cal O}(\epsilon^2) \right\}\;,
\\
  \label{eq:ln}
&L_n = (p_3 \cdot p_4)\,
\frac{2\beta\, }{1+\beta^2}
\int_0^1 \frac{d x}{p(x)^2} \ln^n\left(1+\frac{\vec p_T(x)^2}{p(x)^2}\right) \;,
\\
\label{eq:pn}
&P_n = (p_3 \cdot p_4)\,
\frac{2\beta}{1+\beta^2}
\int_0^1 \frac{d x}{p(x)^2} \text{Li}_n\left(-\frac{\vec p_T(x)^2}{p(x)^2}\right)\; ,
\end{align}
with the auxiliary momentum $p(x)$ defined as $p^\mu(x)=xp_3^\mu+(1-x)p_4^\mu$.
While the expression in Eq.~(\ref{eq:iqq34_result}) is valid in arbitrary kinematics, the analytic results for $L_n$ and $P_n$ has been computed, in Ref.~\cite{Catani:2023tby}, with the assumption of a back-to-back configuration of the final-state emitters. The generalisation of these results thus required to evaluate the one-fold integral representation of $L_n$ and $P_n$ for arbitrary, independent values of final-state momenta.
We performed this step numerically with a Monte Carlo integration.

\subsection{Double soft-gluon emission}
We now consider the double gluon emission contributions. For the initial steps, the computation in Ref.~\cite{Catani:2023tby} is performed for generic values of the hard momenta, and we can thus exactly implement the same strategy, introducing the total soft momenta $k=k_1+k_2$, and performing the $k_1$ and $k_2$ integrations at fixed value of $k$. To this end we introduce the notation
\begin{equation}
\int_{(12)} f(k_1,k_2)\equiv\frac{\Gamma(\frac 12 -\epsilon)}{4^\epsilon \pi^{\frac 12-\epsilon}}\int d^D k_1\,d^D k_2\,f(k_1,k_2)\,\delta_+(k_1^2)\,\delta_+(k_2^2)\,\delta^{(D)}(k-k_1-k_2)\;.
\end{equation}
The relevant integrands are then given by $\int_{(12)} {\cal S}_{ij}$, defined in Eq.~(\ref{eq:ggW}).
This contribution was split, in Ref.~\cite{Catani:2023tby}, into two pieces: 
\begin{equation}
\label{eq:gg_split}
 {\cal S}_{ij}(k_1,k_2)= \frac{p_i^\mu}{p_i\cdot(k_1+k_2)}\, \widetilde\Pi_{\mu\nu}(k_1,k_2)\,\frac{p_j^\nu}{p_j\cdot(k_1+k_2)}+\tilde{\cal S}_{ij}(k_1,k_2)\;,
\end{equation}
with
\begin{equation}
\widetilde\Pi^{\mu\nu}(k_1,k_2) =
- \frac{1}{(k_1\cdot k_2)^2} (-4 g^{\mu\nu}(k_1 \cdot k_2) + (1-\epsilon) k_1^\mu k_2^\nu + (1-\epsilon) k_2^\mu k_1^\nu)\,.
\end{equation}
The first term, proportional to $\widetilde\Pi_{\mu\nu}(k_1,k_2)$, has been isolated since it features a structure similar to the one of the soft current of quark-pair production, which connects their results by the formal substitution
\begin{equation}
n_f T_R\longrightarrow - C_A\, \frac{11-7\epsilon}{4(1-\epsilon)}\;.
\end{equation}  
This relation still holds in the case of general kinematics, allowing us to directly obtain the integral of the first term of Eq.(\ref{eq:gg_split}) by applying this formal substitution to the results of section~\ref{sec:qq}.
The results for the integration of $\tilde{\cal S}$ are instead given by Eqs.~(217) and (218) of Ref.~\cite{Catani:2023tby}.
The back-to-back constraint was imposed in Ref.~\cite{Catani:2023tby} for the leftover integration over $k$, which thus now needs to be performed again for general kinematics.
Since the problem has now effectively been reduced to the integration measure of a single soft momentum, it is analogous to the previously described one-loop single gluon contribution and can be approached in a similar way in $q_T$-space representation.
In this case, however, the momentum $k$ is not forced to be on-shell. The integration measure will thus take the form: 
\begin{align}
\label{eq:2gstructure}
    & \int d^Dk \;\delta^{(D-2)}(\vec q_T+ \vec k_{T}) 
    =\frac {q_T^2}2 \int_0^\infty du\; \frac {dv}{v}
    \;,
\end{align}
with $u= k^2/q_T^2$ and $v = k_-/q_T$, and
where the scalar product with the external momenta $p_3$, $p_4$ can be expressed in terms of the integration variables $u$, $v$ as follows:
\begin{align}
    p_i\cdot k=q_T\left(p_{i+}v+p_{i-}\frac{u+1}{2v}\right)-\vec p_{iT}\cdot\vec q_T\;.
\end{align}
Eq.~(\ref{eq:2gstructure}) eventually leaves us with an additional integration with respect to the single-gluon case. Together with the integration connected with the azimuthal average, we end up with an integration over four variables, to be performed numerically with a Monte Carlo method.

\subsection{NLO-like contributions}
Besides the purely NNLO contributions discussed so far, Eq.~(\ref{eq:h}) depends on the NLO results via their squared average $\braket{(\mathbf{F}_{{\rm ex},1}^{(0)})^2}_{\rm av.}$ and their average squared $\left(\braket{\mathbf{F}_{{\rm ex},1}^{(0)}}_{\rm av.}\right)^2$.

The analytic result for $\braket{\mathbf{F}_{{\rm ex},1}^{(0)}}_{\rm av.}$ valid for general kinematics can be found in Ref.~\cite{Catani:2021cbl}, see Eq.~(22) therein, from which its square can be directly computed. On the other hand, in order to compute the term $\braket{(\mathbf{F}_{{\rm ex},1}^{(0)})^2}_{\rm av.}$ we need to make use of the un-averaged function $\mathbf{F}_{{\rm ex},1}(\phi)$ in general $Q\bar{Q}F$ kinematics. This can also be obtained from the results in Ref.~\cite{Catani:2021cbl}, specifically by combining the results of Eqs.~(22, 25) therein.
The azimuthal average of the squared NLO contribution can then be performed using the relation in Eq.~(\ref{eq:azimuthal average}), integrating numerically over the remaining degrees of freedom.

\subsection{Numerical implementation} 

Our results are made available in a public C++ library named {\sc SHARK} (Soft contributions to Heavy-quark production in ARbitrary Kinematics). 
The library implements all the final results that are available analytically, and all the integrands needed to perform the numerical integrations. The latter are carried out via a Monte Carlo integration, as implemented in the GNU Scientific Library~\cite{galassi2018scientific}.

The final result for the function $\mathbf{h}$ can be decomposed as follows,
\begin{align}
    \mathbf{h} = \; &1 + \frac{\as}{2\pi} \Bigg[ 
    h^{(1)}_{34} \, \textbf{T}_3\cdot\textbf{T}_4 + \sum_{{\substack{I=3,4 \\ j=1,2}}} h_{Ij}^{(1)} \, \textbf{T}_I\cdot\textbf{T}_j
    \Bigg] \nonumber \\
    &+ \left( \frac{\as}{2\pi} \right)^2 \Bigg[ 
    h^{(2)}_{34} \, \textbf{T}_3\cdot\textbf{T}_4 + \sum_{{\substack{I=3,4 \\ j=1,2}}} h_{Ij}^{(2)} \, \textbf{T}_I\cdot\textbf{T}_j
+ h^{(2)}_{123} \, f^{abc} T^a_1 T^b_2 T^c_3 \nonumber
   \\ &+ h^{(2)}_{3434} \textbf{T}_3\cdot\textbf{T}_4 \, \textbf{T}_3\cdot\textbf{T}_4
    + \sum_{{\substack{I=3,4 \\ j=1,2}}}  h^{(2)}_{34Ij} \textbf{T}_3\cdot\textbf{T}_4 \, \textbf{T}_I\cdot\textbf{T}_j
    + \sum_{{\substack{I,K=3,4 \\ j,l=1,2}}}  h^{(2)}_{IjKl} \textbf{T}_I\cdot\textbf{T}_j \, \textbf{T}_K\cdot\textbf{T}_l \Bigg] \nonumber \\ &+ {\cal O}(\as^3)\,.
\end{align}
Our library provides numerical results for all the functions $h$ in the above equation. We note that the decomposition on these particular colour structures is arbitrary, and other decompositions can be obtained which are equivalent upon imposing colour conservation. We also point out that some of the functions multiplying the different colour operators are identical: In particular, the dependence of $h^{(1)}_{Ij}$ reduces solely to $I$, while $h^{(2)}_{IjKl}$ depends only on $I$ and $K$.

As a reference for the user, we provide in Table~\ref{table:benchmarks} the results for the different colour components of $\mathbf{h}$ evaluated at reference points, the latter defined in Table~\ref{table:psp}.
The library SHARK can be downloaded from the public repository: 

{
\centering \href{https://github.com/javimazzitelli/shark}{\texttt{https://github.com/javimazzitelli/shark}}\par
}

\begin{table}
\centering
\renewcommand{\arraystretch}{1.3}
\begin{tabular}{|c c |cccc|}
\hline
 &  & $E$ & $p_x$ & $p_y$ & $p_z$ \\
\hline \hline
\multirow{4}{*}{\rotatebox{90}{Benchmark 1 \hspace{1mm}}} 
& $p_1$ & 195.886 & 0 & 0 & 195.886 \\
& $p_2$ & 195.886 & 0 & 0 & -195.886 \\
& $p_3$ & 97.3393 & 23.2148 & 54.1940 & 77.3076 \\
& $p_4$ & 94.3320 & 34.8122 & -78.5095 & -38.7342 \\
\hline \hline
\multirow{4}{*}{\rotatebox{90}{Benchmark 2 \hspace{1mm}}} 
& $p_1$ & 186.771 & 0 & 0 & 186.771 \\
& $p_2$ & 186.771 & 0 & 0 & -186.771 \\
& $p_3$ & 99.8322 & 96.7478 & 1.61109 & -24.1077 \\
& $p_4$ & 96.0543 & -61.5435 & -25.9394 & 68.8724 \\
\hline \hline
\multirow{4}{*}{\rotatebox{90}{Benchmark 3 \hspace{1mm}}} 
& $p_1$ & 504.806 & 0 & 0 & 504.806 \\
& $p_2$ & 504.806 & 0 & 0 & -504.806 \\
& $p_3$ & 184.499 & 29.6007 & 33.132 & 46.2277 \\
& $p_4$ & 426.944 & -160.136 & -308.706 & 177.227 \\
\hline \hline
\multirow{4}{*}{\rotatebox{90}{Benchmark 4 \hspace{1mm}}} 
& $p_1$ & 715.447 & 0 & 0 & 715.447 \\
& $p_2$ & 715.447 & 0 & 0 & -715.447 \\
& $p_3$ & 307.164 & 151.575 & -123.967 & -161.487 \\
& $p_4$ & 586.261 & -494.56 & 122.074 & -232.983 \\
\hline
\end{tabular}
\caption{Four-momenta of particles $p_1, \dots, p_4$ for four benchmark phase space points. \label{table:psp}}
\end{table}

\begin{table}
\centering
{
\renewcommand{\arraystretch}{1.5}
\begin{tabular}{|c||c||c||c||c|}
\hline
                   & Benchmark 1 & Benchmark 2 & Benchmark 3 & Benchmark 4 \\
                   \hline %\hline
$h^{(1)}_{34}$ & 10.46 & 11.32 & 0.08713 & 0.3289 \\ 
$h^{(1)}_{31}$ & 4.641 & 6.891 & 0.0005135 & 0.09275 \\ 
$h^{(1)}_{41}$ & 6.305 & 5.161 & 0.3844 & 0.7981 \\ 
$h^{(2)}_{34}$ & 39.02(5) & 51.19(5)  & -18.32(2) & -26.70(3) \\
$h^{(2)}_{31}$ & 18.46 & 37.50 & -11.59 & -15.73 \\ 
$h^{(2)}_{41}$ & 29.94 & 22.93 & -11.81 & -14.39 \\ 
$h^{(2)}_{32}$ & 9.820 & 39.47 & -13.63 & -11.06 \\ 
$h^{(2)}_{42}$ & 33.43 & 15.74 & -15.33 & -11.03 \\ 
$h^{(2)}_{123}$ & -4.704(5) & 5.297(5)  & 0.7139(13) & 0.733(3) \\
$h^{(2)}_{3434}$ & 22.94 & 71.80 & -1.295 & -2.606 \\ 
$h^{(2)}_{3431}$ & 31.11 & 73.73 & 0.6334 & 3.351 \\ 
$h^{(2)}_{3441}$ & 21.35 & 82.41 & -5.105 & -9.312 \\ 
$h^{(2)}_{3131}$ & -32.00 & -37.22 & -0.08262 & -1.738 \\ 
$h^{(2)}_{3141}$ & 49.15 & 77.89 & 0.6327 & 3.752 \\ 
$h^{(2)}_{4141}$ & -36.40 & -33.71 & -5.001 & -8.917 \\    
    \hline
\end{tabular}
}
\caption{Values of the different colour components of the function $\mathbf{h}$ evaluated in the four benchmark points defined in Table~\ref{table:psp}. When present, the numbers in parenthesis indicate the uncertainty on the last digit, while when absent the error is lower than the last digit shown. \label{table:benchmarks}}
\end{table}

\section{Summary}
\label{sec:conclusions}

In this work we have computed the ${\cal O}(\as^2)$ soft-parton contributions at low transverse momentum for heavy-quark production in association with a colour singlet. This completes the NNLO expansion of the corresponding transverse-momentum resummation formula, thereby enabling the extension of the NNLO $q_T$-subtraction formalism and the {\sc MiNNLO}$_\text{PS}$ method to $Q\bar{Q}F$ final states.

We provide our results in a public C++ library that implements both the analytic expressions and the numerical evaluation, performed on-the-fly through Monte Carlo integration. These results, already used in existing NNLO and NNLO+PS implementations for $Q\bar{Q}F$ processes, are now publicly available to facilitate further applications.

\section*{Acknowledgments}
We are grateful to Massimiliano Grazzini for his support on this project and for providing feedback on the manuscript. We thank Luca Buonocore, Pier Monni and Luca Rottoli for useful cross-checks.
The work of S.D. has been funded by the European Union (ERC, MultiScaleAmp, Grant Agreement No. 101078449). Views and opinions expressed are however those of the author(s) only and do not necessarily reflect those of the European Union or the European Research Council Executive Agency. Neither the European Union nor the granting authority can be held responsible for them.
\bibliography{long}
\bibliographystyle{JHEP}

\end{document}